\documentclass{iopart}

\usepackage{iopams}  
\usepackage{graphicx}

\begin{document}

\title[Effect of interface disorder on QW excitons and MC
polaritons]{Effect of interface disorder on quantum well excitons and microcavity
polaritons}

\author{Vincenzo Savona}

\address{Institut de Th\'eorie des Ph\'enom\`enes Physiques, Ecole
Polytechnique F\'ed\'erale de Lausanne (EPFL), CH-1015 Lausanne, Switzerland}
\ead{vincenzo.savona@epfl.ch}

\begin{abstract}
The theory of the linear optical response of excitons in quantum wells and polaritons in planar semiconductor microcavities is reviewed, in the light of the existing experiments. For quantum well excitons, it is shown that disorder mainly affects the exciton center-of-mass motion and is modeled by an effective Schr\"odinger equation in two dimensions. For polaritons, a unified model accounting for quantum well roughness and fluctuations of the microcavity thickness is developed. Numerical results confirm that polaritons are mostly affected by disorder acting on the photon component, thus confirming existing studies on the influence of exciton disorder. The polariton localization length is estimated to be in the few-$\mu$m range, depending on the amplitude of disorder, in agreement with recent experimental findings.
\end{abstract}

\pacs{68.35.Ct, 71.35.-y, 71.36.+c, 73.21.-b}
\maketitle

\section{Introduction}

Planar semiconductor heterostructures are unavoidably characterized by structural defects arising in the fabrication process. This disorder affects the optical response of the lowest-lying excited states, particularly the exciton. Although initially
disorder was considered an unwanted feature \cite{Weisbuch1981}, in the light of the study of ideal systems
of lower dimensionality, later it was realized that it might induce new behaviors and
essentially open the way to the study of fundamental effects and to possible applications \cite{Zimmermann2003}. In particular, disorder breaks the in-plane translational symmetry and can lead to a dramatic change in the physical behaviour of the interband excitations, producing multiple scattering and localization. This behaviour has several implications in the optical response of the system. In particular, it produces an inhomogeneous spectral broadening that is often used as a figure of merit of the interface quality. 

Disorder in heterostructures and its influence on the exciton optical response have been a topic of intense research in the last two decades \cite{Zimmermann2003}. In addition to inhomogeneous spectral broadening, in fact, disorder in heterostructures leads to several manifestations of fundamental quantum physics of disorder, that have been over the years one of the main motivation for these studies. In quantum wells (QWs) in particular, the exciton motion subject to weak disorder is an almost ideal physical realization of a particle obeying Schr\"odinger equation in presence of a two-dimensional static disorder potential. Its optical response provides direct access to the eigenstates of the exciton motion, and is an excellent mean for studying the physics of Anderson localization. Apart from the fundamental implications of these studies, a more practical motivation resides in the possibility of an accurate characterization of the interface structure, giving access to information complementary to that obtained by means of microscopy techniques. 

More recently, the influence of heterointerface disorder on exciton-polaritons in semiconductor microcavities (MCs) was also investigated \cite{Ell1998,Gurioli2001,Langbein2002a,Langbein2004a,Litinskaia2001,Litinskaia2000,Houdre2000,Michetti2005,Savona1997a,Whittaker1998,Whittaker1996}. In this system, the fundamental excitations are the two-dimensional analog of bulk polaritons, namely normal modes of the linear coupling between excitons in a QW and the electromagnetic mode of a planar resonator \cite{Kavokin2003d,Khitrova1999,Savona1999}. Most of the existing theoretical studies have concentrated on the influence of QW disorder on polaritons \cite{Litinskaia2001,Litinskaia2000,Savona1997a,Whittaker1998,Whittaker1996}, with the conclusion that polaritons are almost unaffected by QW disorder provided that the amplitude of the energy fluctuations is sufficiently smaller than the Rabi energy characterizing the linear exciton-photon coupling. Some experiments \cite{Gurioli2001,Houdre2000,Langbein2002a,Langbein2004a} have recently suggested that disorder at the MC interfaces might affect the polariton quasi-particle even more than QW disorder, through its influence on the spectral properties of the resonant electromagnetic mode. In particular, unambiguous signatures of long-range defects characterizing the MC interfaces appear in the angular pattern of resonantly scattered light. These measurements even suggest that polaritons can undergo localization in the same way as excitons, however over a much larger spatial range of the order of a few $\mu$m. 

There have been many theoretical studies of the influence of QW disorder on the optical response of excitons. These studies have been recently reviewed by Zimmermann {\em et al.} \cite{Zimmermann2003} in a very comprehensive manner. The influence of disorder on polaritons has received less attention in the past. In particular, theoretical studies have principally addressed the effect of QW disorder while completely overlooking the role of disorder acting on the photon component of polaritons \cite{Litinskaia2001,Litinskaia2000,Savona1997a,Whittaker1998,Whittaker1996}.

Here, I will give an overview of the influence of disorder on excitons and polaritons in heterostructures. In the first part, I will briefly review the basic properties of QW excitons in presence of weak disorder. This review is only meant to introduce the main physical aspects that are later important in the context of MC polaritons. A more comprehensive review is the one by Zimmermann {\em et al.} \cite{Zimmermann2003}. In the second part, I will discuss the present status of the research on the effect of disorder on MC polaritons. In this context, I will present some original results stemming from the numerical analysis of a model that accounts both for the exciton and the photon components of disorder. This preliminary analysis clarifies the mutual roles of exciton and photon disorder, and confirms the first experimental evidence that photon disorder has the strongest influence on polaritons. It also provides a first estimate of the polariton localization length, which ranges from a few $\mu$m to a few tens of $\mu$m in typical samples. 

\section{Quantum well excitons}

Disorder in QWs has mainly two origins. First, the height fluctuations of the QW interfaces, resulting in a variation of the QW thickness. Second, alloy disorder, arising when a semiconductor alloy is used but also at an interface between two pure species (e.g. GaAs-AlAs) due to segregation or interdiffusion of atoms at the interface \cite{Zimmermann1997}. Both effects result in an in-plane fluctuation of the electron and hole confinement energy in the QW. While alloy disorder is short-ranged, interface fluctuations can take place over a correlation length of 10 to 100 nm, especially if growth interruption is used \cite{Savona2006}. Here, we are mainly interested in how QW disorder affects the exciton states. QW excitons display inhomogeneous spectral broadening that was shown by Weisbuch {\em et al.} \cite{Weisbuch1981} to be related to the thickness fluctuations resulting in a variation of the exciton confinement energy in the QW. By that time it was still unclear, however, that disorder can affect not only the potential but also the kinetic energy of the exciton moving along the QW plane. The exciton then behaves as a massive particle subject to a disordered potential, giving rise to spatially localized eigenstates of the center of mass (COM).

The theory of exciton states and their optical properties in disordered QWs has been mainly developed over the last two decades by R. Zimmermann. A complete review of this topic is found in Ref. \cite{Zimmermann2003}. Here we present a short overview of the basic elements of the theory, highlighting the aspects that are most relevant to the MC polariton problem, that is addressed in the next section.

In principle, an exciton in a disordered QW is described, within the envelope function approximation, by a two-particle Schr\"odinger equation for the electron and the hole in three dimensions
\begin{eqnarray}  \label{ExSchroe}
\lefteqn{\bigg ( - \frac{\hbar^2}{2m_e} \Delta_{{\bf r}_e}
    - \frac{\hbar^2}{2m_h} \Delta_{{\bf r}_h}
    - \frac{e^2/\epsilon_0}{|{\bf r}_e - {\bf r}_h|}}&&\nonumber\\
&\qquad\qquad&+ W_e({\bf r}_e) + W_h({\bf r}_h) - \epsilon_\alpha \bigg ) \Psi_\alpha ({\bf r}_e,{\bf r}_h) = 0 \, .
\end{eqnarray}
The confinement potentials $W_a({\bf r}_a) \; (a=e,h)$
describe the spatial variation of the local band edges. The $z$-axis
is taken along the growth direction. These potentials depend on the band edge difference between barrier and well material, and contain all information on the QW disorder. In most QWs of average quality, both based on III-V and II-VI materials, the exciton binding energy is significantly smaller than the inhomogeneous spectral broadening. This suggests that disorder has only a weak influence on the electron-hole relative motion in the exciton state. A more rigorous formulation of this idea can be given in terms of the {\em rigid exciton approximation}. In practice, it is assumed that the perturbation introduced by disorder is not sufficient to produce a transition from the exciton $1s$ state to higher states of the relative electron-hole motion. Hence, the exciton stays always in the $1s$ state and only its COM motion is affected by disorder. We make the following factorization ansatz for the exciton wave function
\begin{equation}          \label{FactorWF}
\Psi_\alpha ({\bf r}_e,{\bf r}_h) =
  f_e(z_e) \, f_h(z_h) \, \phi_{1s}({\pmb \rho}_e-{\pmb \rho}_h) \, \psi_\alpha({\bf R}) \, ,
\end{equation}
where ${\bf R} = (m_e{\pmb \rho}_e + m_h{\pmb \rho}_h)/M$ is the COM coordinate in two dimensions and $M=m_e+m_h$ the exciton kinetic mass. Here we take $f_a(z_a)$ and $\phi_{1s}({\pmb\rho})$ as the solutions of the Schr\"odinger equation of an ideally flat QW structure having a confinement potential $W_a(z)=\overline{W_a({\pmb \rho},z)}$ averaged over the in-plane coordinate. The factorized ansatz (\ref{FactorWF}) is an exact solution of the Schr\"odinger equation with the translation-invariant potential $W_a(z)$. We can therefore perform a first-order perturbation theory with respect to the difference $W_a({\bf R},z) - W_a(z)$. This results in an effective Schr\"odinger equation for the COM wave function
\begin{equation}            \label{COM}
\left( - \frac{\hbar^2}{2M} \nabla^2 + V_x({\bf R}) \right) \, \psi_\alpha({\bf R}) =
\epsilon_\alpha \, \psi_\alpha({\bf R}) \, ,
\end{equation}
where the $1s$ exciton energy $\hbar\omega_x$ of the averaged QW has been set to zero for simplicity. The normalization relation for the COM wave functions over the quantization area $A$ reads
\begin{equation}             \label{Ortho}
\int_{A} d{\bf R} \, \psi_\alpha({\bf R}) \, \psi_\beta({\bf R}) = \delta_{\alpha,\beta} \, .
\end{equation}
The effective COM potential $V_x({\bf R})$ resulting from this approach is
\begin{equation}            \label{VR}
V_x({\bf R}) =
  \int \! d{\bf R}^\prime \sum_{a=e,h} \eta_a^2 \, \phi_{1s}^2(\eta_a({\bf R}-{\bf R}^\prime)) \,
 U_a({\bf R}^\prime) \, ,
\end{equation}
where $U_a({\bf R}^\prime)$ represents the electron and hole confinement-energy fluctuations along the QW plane. This result shows that the effective COM potential derives from the actual energy fluctuation averaged over the relative e-h motion. Due to the mass factors $\eta_e = M/m_h, \, \eta_h = M/m_e$, different weights
are given to the electron and the hole contribution. In materials such as GaAs/AlGaAs, for example, the hole mass is larger and the well width fluctuations are averaged less efficiently, making the hole contribution dominant in (\ref{VR}). The validity of this rigid exciton approximation has been checked for typical QWs made of III-V semiconductors \cite{Zimmermann1997} and turns out to provide an extremely good description of the exciton optical response when the inhomogeneous exciton broadening does not exceed the $1s$ binding energy. 

A microscopic model of the exciton disorder potential would require a detailed knowledge of the interface structure responsible of the confinement energy fluctuation $U({\bf R})$. This structure depends dramatically on the growth conditions. For high quality structures made using growth interruption at both QW interfaces \cite{Savona2006}, for example, interfaces are characterized by monolayer steps having size of up to 100 nm and anisotropic shape with a preferential orientation \cite{Zimmermann2001,Gammon1996}. In this case, a very detailed model for $V_x({\bf R})$ can be built on the basis of the experimental observations \cite{Savona2006}. In most situations in which growth interruption is not applied, on the other hand, the correlation length of interface fluctuations is expected to be shorter. As Eq. (\ref{VR}) suggests, the short-range details of the fluctuating potential $U({\bf R})$ are smeared out in the convolution with the $1s$ exciton wave function, which extends over a distance given by the exciton Bohr radius. Then, a reliable model for the effective COM potential $V_x({\bf R})$ is given in terms of a Gauss-distributed spatially-correlated random potential characterized by the correlation 
\begin{equation}
\langle V_x({\bf R})V_x({\bf R}^\prime)\rangle=\sigma_x^2f(|{\bf R}-{\bf R}^\prime|)\,,
\label{potcorr}
\end{equation}
where $\sigma_x$ is the standard deviation of the Gauss energy distribution and $f(R)$ is a correlation function equal to 1 at $R=0$ and decaying to zero over a correlation length $\xi_x$. This simple disorder model was successful in describing several features of the exciton optical response, including the level-repulsion in the statistics of energy level distances measured by near-field optical spectroscopy \cite{Intonti2001,Lienau2004}.

\begin{figure}
\centerline{\includegraphics[width=5.0in]{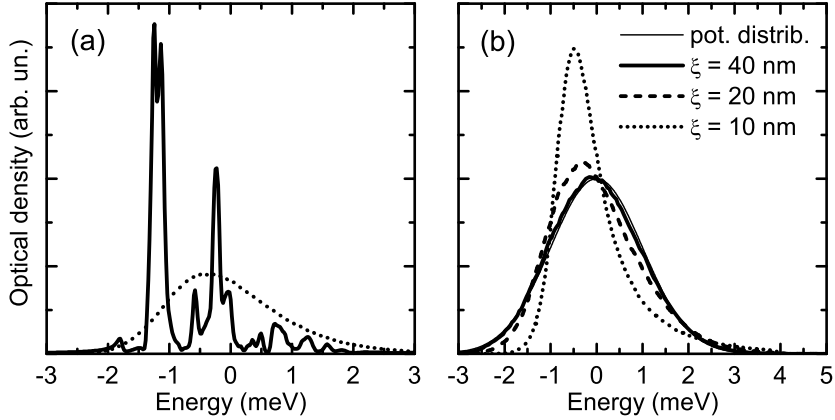}}
\caption{(a): Simulated optical density for a single potential realization (solid line) and averaged over $10^4$ realizations (dotted line). The simulation parameters are $\sigma_x=1$ meV and $\xi_x=10$ nm. For the single realization spectrum, a homogeneous linewidth $\hbar\gamma=50~\mu\mbox{eV}$ has been introduced. (b): Simulated optical density for different values of the potential correlation length $\xi_x$ and $\sigma_x=1~\mbox{meV}$.}
\label{fig1} 
\end{figure}

We now turn to the problem of modeling the optical response of the eigenstates of Eq. (\ref{COM})
For the evaluation of the optical response we derive the optical matrix element for the transition between the semiconductor ground state and the exciton state (\ref{FactorWF}). The dipole Hamiltonian reads
\begin{equation}        \label{DipHam}
H_{dip}=-\frac{e}{mc}\sum_j {\mathbf A}({\mathbf r}_j)\cdot {\mathbf p}_j \, ,
\end{equation}
where the sum runs over all the electrons, ${\mathbf A}({\mathbf r})=A^{(0)}{\mbox{\boldmath $\epsilon$}}\exp[i(k_zz+{\mathbf k}\cdot{\mathbf R})]$ is the vector potential for a plane-wave electromagnetic field, and ${\pmb \epsilon}$ is the light polarization direction. Within the effective-mass and envelope-function approximations, the matrix element for the optical transition can be shown to result in
\begin{equation}       \label{DIP}
\langle\Psi_\alpha|H_{dip}|0\rangle={\mbox{\boldmath $\epsilon$}}\cdot{\mbox{\boldmath $\mu$}}_{cv}\phi_{1s}(0)O_{eh}(k_z)\int_A d{\mathbf R} \, e^{i{\mathbf k}\cdot{\mathbf R}}\, \psi_\alpha(\mathbf{R}) \, ,
\end{equation}
with the Bloch part of the matrix element given by the integral over the Brillouin zone
\begin{equation}       \label{Mucv}
{\mbox{\boldmath $\mu$}}_{cv}=\int_{BZ}d{\mathbf r} \, u^*_v({\mathbf r})\, e\, {\mathbf r} \, u_c({\mathbf r}) \, ,
\end{equation}
and the z-confinement overlap integral
\begin{equation}         \label{Oeh}
O_{eh}(k_z)=\int dz \, f_e(z)\, f_h(z)\, e^{ik_zz} \, .
\end{equation}
This latter quantity is practically equal to unity for all typical QW structures and is therefore omitted in what follows.
The optical density is obtained from the Fermi golden rule
\begin{equation}          \label{ODFermi}
D(\omega)=\frac{2\pi}{\hbar}\sum_\alpha |\langle\Psi_\alpha|H_{dip}|0\rangle|^2\delta(\omega-\omega_\alpha) \, .
\end{equation}
By omitting $\alpha$-independent prefactors and introducing the Fourier-transformed wave function
\begin{equation}
\psi_{\alpha {\mathbf k}} = \int_A d{\mathbf R} \, e^{i{\mathbf k} \cdot {\mathbf R}} \, \psi_\alpha(\mathbf{R}) \, ,
\end{equation}
we can rewrite the optical density by weighting with the
wave function squared at ${\mathbf k} = 0$,
\begin{equation} \label{OD}
D(\omega) = \frac{1}{A} \sum_\alpha \psi^2_{\alpha {\mathbf k}=0 } \, \delta(\omega - \omega_\alpha) \, .
\end{equation}
In the context of the COM problem (\ref{COM}), this is
the spectral function at zero momentum, $D(\omega) = \Im G_{{\mathbf k}=0}(\omega - i0^+)$,
where $G_{\mathbf k}(y)$ is the one-particle (disorder-averaged) Green's function.
More important, $D(\omega)$ is proportional to the absorption lineshape of
1s excitons at normal incidence.
In a given potential realization, expression (\ref{OD})
consist of a series of spectral lines that form the exciton inhomogeneous spectrum (also accounting for a small homogeneous broadening of each  spectral line). Experimentally, a small angular average in the detection process is always introduced. It can be shown \cite{Savona1999b} that this average is simulated by performing a statistical average of the spectrum over a large number of random realizations of the disorder potential $V_x({\bf R})$. This produces smooth curves, as shown in the simulation in Fig. \ref{fig1}(a). 

The simulated optical density points out to the role of quantum mechanics in determining the exciton COM spectral properties. By rewriting the Schr\"odinger equation in a dimensionless fashion \cite{Savona1999b}, the shape of the COM spectral properties depends on a single dimensionless parameter $\sigma_x/E_{cx}$, where $E_{cx}=\hbar^2/(2M\xi_x^2)$ is a characteristic confinement energy, given the potential amplitude $\sigma_x$ and its correlation length $\xi_x$. In the limit $\sigma_x/E_{cx}\rightarrow\infty$ the kinetic term in the Schr\"odinger equation vanishes compared to the potential term. Then, all the COM eigenstates $\psi_\alpha({\bf R})$ tend to Dirac-delta functions in real space, and only the disorder potential determines the particle spectral function. This is the classical limit and corresponds to the interpretation given by Weisbuch et al. \cite{Weisbuch1981} for the dependence of the exciton inhomogeneous broadening on the average QW thickness. For smaller $\sigma_x/E_{cx}$ the kinetic term becomes important and the spatial extension of the COM eigenstates $\psi_\alpha({\bf R})$ increases, implying a further spatial averaging of the disorder potential $V_x({\bf R})$. As a result the particle spectrum becomes narrower and asymmetric -- an effect that has been traditionally called {\em motional narrowing} \cite{Zimmermann1997}. The dependence of the optical density spectrum on the dimensionless parameter $\sigma_x/E_{cx}$ is illustrated in Fig. \ref{fig1}(b). A scaling argument can be formulated \cite{Savona1999b}, according to which the inhomogeneous broadening $\hbar\Gamma_{inh}$ in the quantum limit $\sigma_x/E_{cx}\ll1$ scales as $\hbar\Gamma_{inh}\approx\sigma_x(\sigma_x/E_{cx})$. Hence, in this extreme limit, the influence of disorder on the particle motion can become vanishingly small. The concept of motional narrowing has been the subject of intense debate in connection to the system of MC polaritons, as we will discuss in detail below. In standard QWs based on III-V semiconductors, motional narrowing is of considerable importance. For example, if $\sigma_x=0.3$ meV and the correlation length $\xi_x=10$ nm, then $\sigma_x/E_{cx}\approx0.3$.

\begin{figure}
\centerline{\includegraphics[width=4in]{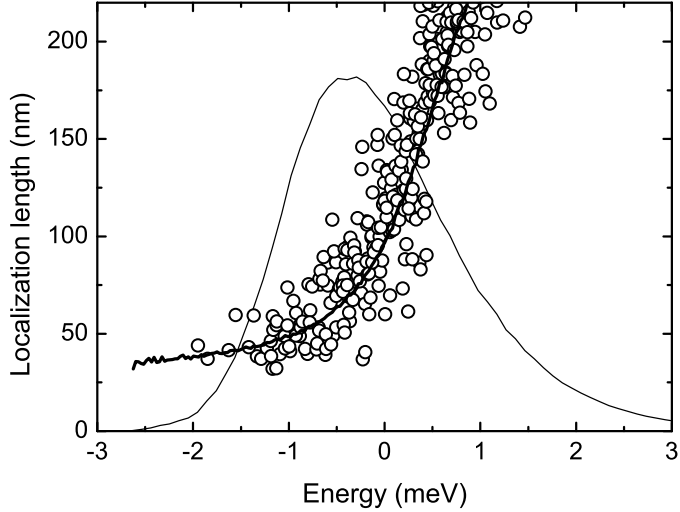}}
\caption{Exciton COM localization length computed according to Eq. (\ref{IPN}). Simulation parameters as in Fig. \ref{fig1} (a). The circles are values for specific COM eigenstates extracted from simulations of a few disorder realizations. The thick line represents the COM localization length averaged over $10^4$ disorder realizations. The averaged optical density is plotted as a thin line for reference.}
\label{fig2} 
\end{figure}

A good estimate of the spatial extension of the localized COM states can be obtained via the {\em participation ratio} \cite{Kramer1993}. The exciton localization length $\Lambda_{\alpha}$ of the state $\psi_\alpha(\mathbf{R})$ is defined as
\begin{equation}              \label{IPN}
\Lambda_{\alpha}^{-d}=\int d{\mathbf R} \, \psi_\alpha^4({\mathbf r}) \, ,
\end{equation}
where $d$ denotes the dimensionality of the system. The fourth power of the wave function weights the regions where the amplitude is large (the same integral with exponent two is equal to one because of wave function normalization). We remind that, according to Anderson's scaling argument \cite{Abrahams1979}, all wave functions in a two-dimensional system are expected to be localized. The particle localization length is distributed over a broad range even for a fixed energy, as the sample calculation in Fig. \ref{fig2} shows. Its energy-dependent average displays a dramatic rise as a function of energy, across the exciton line. As the localization length becomes larger than the exciton coherence length, this latter then governs the exciton motion that becomes diffusive. This explains the occurrence of an exciton {\em mobility edge}, which has been object of several investigations in the early experiments on QWs \cite{Hegarty1984,Jahn1997}. As for the inhomogeneous broadening, also the particle localization length can be deduced from a scaling argument. In two dimensions and in the quantum limit $\sigma_x/E_{cx}\ll1$ in particular, and the localization length must scale as $\Lambda_{\alpha}\approx\xi_x(E_{cx}/\sigma_x)$, thus approaching the limit of Anderson localization $\Lambda_{\alpha}\gg\xi_x$.

When studying the optical response of excitons in QWs with rough interfaces, the most important effect of disorder is the Resonant Rayleigh scattering (RRS) of light. RRS can be seen as the sum of the electromagnetic field scattered by all the localized exciton states when excited resonantly by a light beam. The RRS signal originating from a plane-wave excitation is emitted in all directions, due to the breaking of in-plane momentum conservation. It is a coherent scattering, hence it is characterized by speckles, namely intensity fluctuations over frequency or emission angle that arise from multiple interference of light coming from all exciton scatterers. The very first experimental studies of RRS \cite{Hegarty1982,Stolz1993,Stolz1994} were aimed at measuring the exciton homogeneous linewidth that characterizes the RRS spectrum under monochromatic excitation. Later, the properties of RRS were studied in more detail, showing that its time-dependence could be related to the microscopic nature of the localized exciton states \cite{Zimmermann2003}. In the last decade, in particular, RRS from excitons in QWs was the object of intense experimental \cite{Gurioli1996,Garro1997,Haacke1997,Marie1997,Gurioli1997,Hayes1998,Birkedal1998,Woerner1998,Langbein1999,Garro1999,Hayes2000,Haacke2000,Langbein2000,Prineas2000,Kocherscheidt2003,Schwedt2003} and theoretical \cite{Zimmermann1995,Belitsky1995,Citrin1996,Savona1999b,Malpuech2000,Prineas2000} investigation. The RRS signal could be isolated from the contribution of incoherent photoluminescence to the total light emission into non-specular directions (secondary emission) \cite{Haacke1997,Hayes1998}, in particular by means of spectral interferometry \cite{Birkedal1998,Gurioli1997,Hayes2000,Haacke2000} and statistical analysis of the speckle pattern \cite{Langbein1999,Langbein2000}. It was also shown that time-resolved RRS can bring direct evidence for quantum-mechanical energy-level repulsion \cite{Savona1999b,Kocherscheidt2003,Savona2000a,Malpuech2000}. Recently, time-resolved RRS was also used together with theoretical modeling as an efficient tool for studying the details the QW interface structure \cite{Savona2006}.

Excitons in disordered QWs represent a typical example of a quantum particle moving in a disordered landscape, governed by the Schr\"odinger equation. It is then possible to investigate many aspects of the fundamental physics of disorder and localization. Anderson localization \cite{Kramer1993} was actually observed in near-field measurements \cite{freymann:394,Gammon1996,Intonti2001,Lienau2004}, thanks to the fact that an optical measurement can probe the bottom of the disorder energy band, as opposite to transport measurements in metals, where the effect of disorder on electrons close to the Fermi energy only can be accessed. Recently, in particular, two fundamental phenomena have been predicted and successively measured in QWs: the statistical distribution of energy levels, leading to {\em level repulsion}  \cite{Intonti2001,Lienau2004,Savona1999b,Feltrin2004}, and the enhanced resonant backscattering \cite{Langbein2002b,Savona2000}. This phenomenon consists in an enhancement of the average scattered intensity within a cone centered at the backscattered direction ${\bf k}=-{\bf k}_{in}$. It originates from the constructive interference between each multiple scattering path and its time-reversed inside the scattering medium \cite{Akkermans1986,Albada1985,Kramer1993,Wiersma1997,Wolf1985}. As such, it cannot be accounted for by a finite-order perturbation theory and is considered as a precursor of Anderson localization. In a system showing localization, in fact, the angular width of the backscattering peak is a measure of the localization length. Enhanced backscattering was recently predicted and measured for excitons in very high quality QWs \cite{Langbein2002b,Savona2000}.

\section{Microcavity polaritons}

\begin{figure}[h!]
\centerline{\includegraphics[width=3.0in]{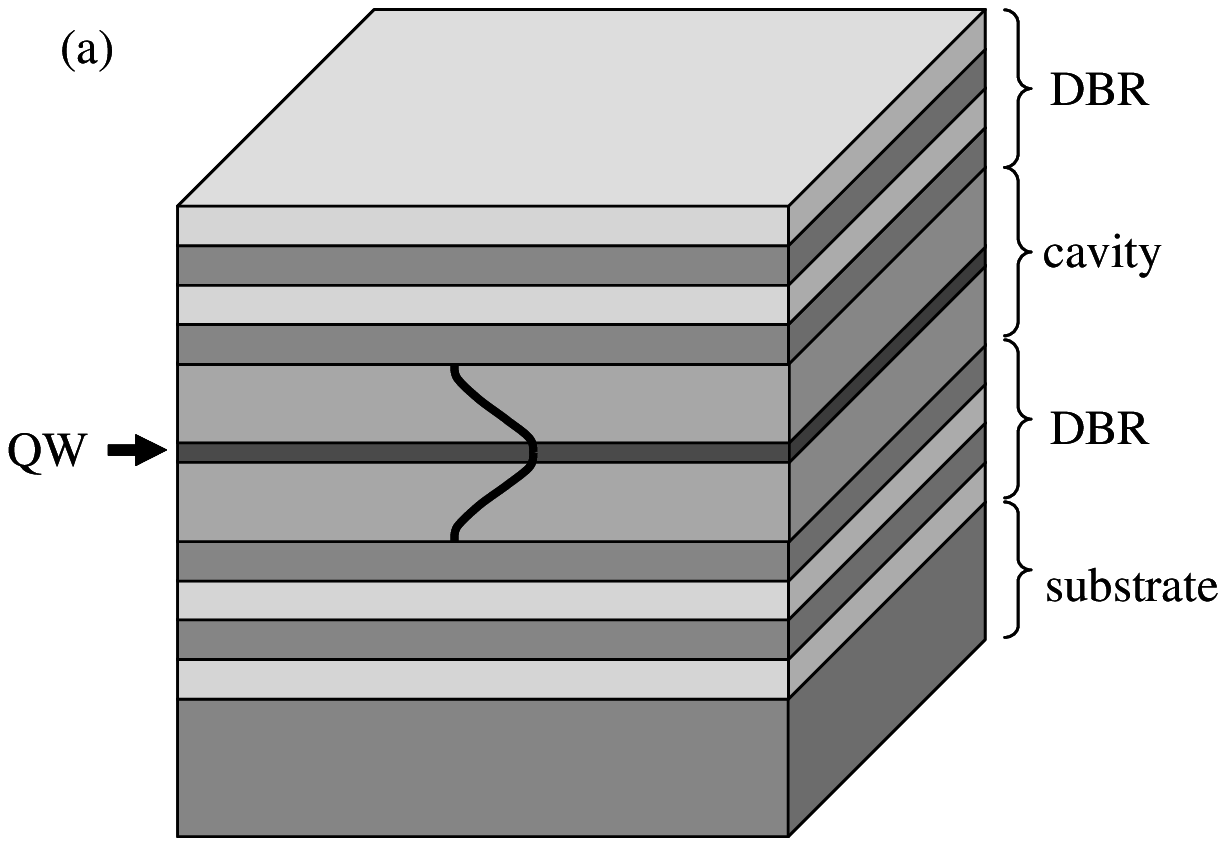}\includegraphics[width=2.5in]{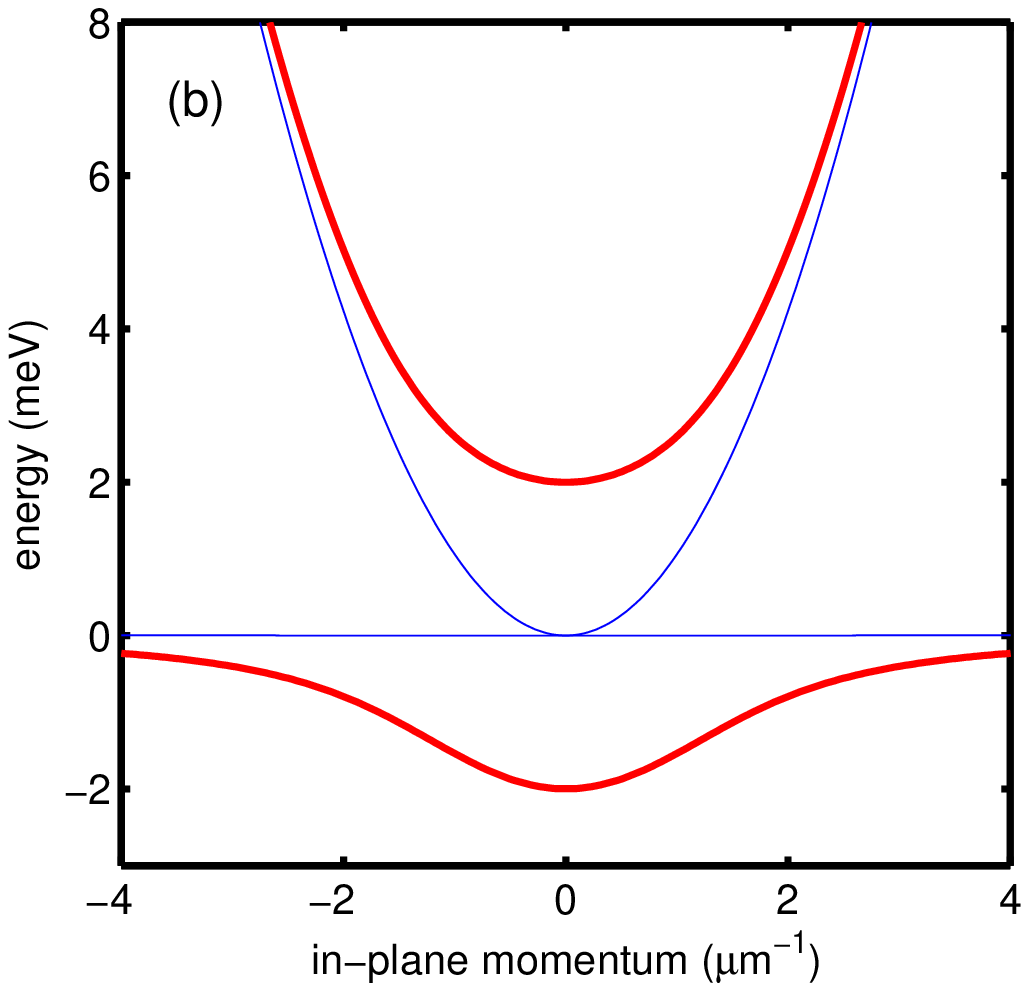}}
\caption{(a) Schematic structure of a semiconductor MC embedding a QW. The number of layers in the DBRs as well as the QW thickness are not on scale. The shape of the resonant electromagnetic mode in the cavity layer is sketched. (b) Energy-momentum dispersion curves of the polariton modes obtained from (\ref{polham}) assuming zero detuning and $\Omega=2$ meV. The thin (blue) lines are the uncoupled exciton and photon energies $E_{x\bf k}$ and $E_{p\bf k}$.}
\label{fig01pol}
\end{figure}

MC polaritons arise from the normal-mode coupling between an exciton in a QW and the lowest-frequency electromagnetic mode of a Fabry-P\'erot planar resonator in which the QW is embedded \cite{Kavokin2003d,Khitrova1999,Savona1999}. For an ideal system without disorder, the in-plane translational symmetry implies that the in-plane momentum ${\bf k}$ is a good quantum number. Hence, one exciton with a given ${\bf k}$ is linearly coupled to one photon mode having the same ${\bf k}$. The simplest way of modeling the system \cite{Savona1995} is in terms of a ${\bf k}$-dependent eigenvalue problem 
\begin{equation}
H_{\bf k}
\left(\begin{array}{c}
\psi_{p\bf k}\\
\psi_{x\bf k}
\end{array}\right)
=E_{\bf k}
\left(\begin{array}{c}
\psi_{p\bf k}\\
\psi_{x\bf k}
\end{array}\right)\,,
\label{poleq}
\end{equation}
where
\begin{equation}
H_{\bf k}=\left(
\begin{array}{cc}
E_{p\bf k}-i\hbar\gamma_{p\bf k} & \displaystyle \frac{\hbar\Omega_{\bf k}}{2}\\
\displaystyle \frac{\hbar\Omega_{\bf k}}{2} & E_{x\bf k}-i\hbar\gamma_{x\bf k}
\end{array}\right)\,.
\label{polham}
\end{equation}
Here, ${\bf k}$-dependent exciton and photon damping rates $\gamma_{x\bf k}$, $\gamma_{p\bf k}$ were included. The rates model, respectively, the photon escape out of the cavity and the possible exciton non-radiative dissipation mechanisms. The quantities $E_{p\bf k}$ and $E_{x\bf k}$ are the photon and exciton energy-momentum dispersion relations, while $\hbar\Omega_{\bf k}$ is the linear exciton-radiation coupling. Typically, ${\bf k}$-independent coupling constant $\hbar\Omega$ and damping rates $\gamma_{p}$ and $\gamma_{x}$ are assumed. Simple analysis of this eigenvalue problem \cite{Savona1995} indicates that, provided the damping rates are sufficiently smaller than $\hbar\Omega$, {\em strong coupling} arises, with normal modes being linear superpositions of exciton and photon modes, called {\em polaritons}. The normal-mode energy splitting, otherwise named {\em vacuum-field Rabi splitting}, is equal to $\hbar\Omega$ at the resonance condition $E^{x}_0=E^{p}_0$. Fig. \ref{fig01pol}(a) illustrates the structure of a semiconductor MC, highlighting the multilayered mirrors called Distributed Bragg Reflectors (DBRs). Fig \ref{fig01pol} (b) shows the upper and lower polariton dispersion curves, obtained from the diagonalization of (\ref{polham}) at zero exciton-cavity detuning. 

This simple representation is very effective in many experimental situations, mainly as it reproduces very well the measured polariton energies \cite{Houdre1994b,Langbein2004a,Savona1996,Weisbuch1992}. Although it assumes perfectly two-dimensional photon modes, a full three-dimensional treatment of the leakage through the DBRs \cite{Jorda1995,Savona1995a,Savona1996} gives a microscopic account of the effective photon damping rate $\gamma_{p}$. Still, the effective model contains several simplifying assumptions that essentially limit its predictivity. First, it does not account for the cavity {\em leaky modes}, that can be resonant with the exciton band at finite momentum, modifying the polariton dispersion and altering the polariton radiative rate \cite{Savona1995a,Savona1996}. Second, the polariton non-radiative damping rates have to be modeled consistently with the steep polariton dispersion, resulting in a suppression of the damping in the strong-coupling region \cite{Ciuti1998,Savona1997,Tassone1997}. 

The most important limitation of the coupled oscillator model, however, is its inability to model disorder effects -- basically because it assumes in-plane momentum conservation that is instead lifted in presence of disorder. Historically, the problem was brought up in an attempt to explain the measured inhomogeneous broadening of the MC polariton spectral lines and their dependence on exciton-cavity detuning \cite{Whittaker1996}. In particular, the coupled oscillator model (\ref{poleq}), only accounting for homogeneous broadening, predicts equal linewidhs for upper and lower polariton at resonance, given by the average $\hbar(\gamma_{p}+\gamma_{x})/2$. Experiments instead show that, at zero detuning, the lower polariton always features a spectral line narrower than that of the upper polariton. The early theoretical models of polariton inhomogeneous broadening assumed phenomenologically an energy distribution of exciton resonances, linearly coupled to the cavity mode \cite{Andreani1998,Houdre1996,Savona1999,Savona1996a}. This approach can explain how only a small part of the exciton spectral density concurs to the formation of polaritons, while the remaining part is weakly coupled to light and appears as a weak spectral feature at the bare exciton energy. In was then clear that polariton lines can be much narrower than the bare exciton line. In particular, in the limit when $\hbar\Omega\gg\sigma_x\gg\hbar\gamma_x$  the polariton linewidths at zero detuning approach the homogeneous value $\hbar(\gamma_{p}+\gamma_{x})/2$ \cite{Houdre1996}. Still, however, this model predicts equal upper and lower polariton linewidths at resonance. 

We will discuss below how this {\em linear dispersion model} can be  more rigorously derived from a microscopic theory in presence of disorder. A very complete account of the implications of the linear dispersion model is given in the review article by Khitrova {\em et al.} \cite{Khitrova1999}. Before proceeding, let us briefly remind that the linear dispersion model was originally introduced by Zhu {\em et al.} \cite{Zhu1990} in connection to the atom-cavity vacuum-field Rabi splitting in atomic physics. In the atom-cavity case, in fact, it was not clear whether the vacuum-field Rabi splitting, obtained when a single atom is coupled to the cavity mode, was a purely quantum effect. When $N$ atoms are present in the cavity, instead, the Rabi splitting increases as $\sqrt{N}$ \cite{Kaluzny1983,Thompson1992}. This increase is indeed a quantum effect originating from the nonlinearity in the optical response of a two-level system and modeled by the Jaynes-Commings Hamiltonian \cite{Jaynes1963}. Zhu {\em et al.} proved that in the single-atom case instead, the vacuum-field Rabi splitting was not a quantum effect and could simply be explained in terms of the classical response of a Lorentz oscillator. The debate on the classical vs quantum origin of the Rabi splitting was also started in connection to MC polaritons shortly after their first observation by Weisbuch {\em et al.} \cite{Weisbuch1992}, and is discussed extensively in the review by Khitrova {\em et al.} \cite{Khitrova1999}

\subsection{Influence of exciton disorder}

The effect of disorder on polaritons was first studied in connection to the interpretation of inhomogeneous polariton spectra, both experimentally \cite{Whittaker1996,Bongiovanni1997} and theoretically \cite{Whittaker1996,Savona1997a,Whittaker1998}. All these studies assumed that only disorder of the QW interfaces affects polaritons through their exciton component. As already pointed out, experiments brought systematic evidence of a lower polariton line narrower than the upper polariton one. 

On the theoretical side, Whittaker {\em et al.} \cite{Whittaker1996} were the first to study the problem in terms of in-plane propagation of the exciton subject to a disorder potential acting on its COM. They suggested that {\em polariton motional narrowing} could explain the small inhomogeneous broadening of the lower polariton at resonance. This conclusion however stems from the assumption that the polariton is a quasi-particle with a perfect parabolic dispersion, determined by its effective mass, at all momenta. In particular, in Ref. \cite{Whittaker1996} this assumption was used to evaluate the polariton linewidth from a scaling argument, that was however applied incorrectly. Correct use of the scaling argument \cite{Whittaker1998,Litinskaia2000,Litinskaia2001} predicts a polariton inhomogeneous broadening $\hbar\Gamma_p$ scaling as $\sigma_x^2/E_{cx}$. The correlation energy is here defined as $E_{cp}=\hbar^2/(2M_p\xi_x^2)$ in terms of the polariton effective mass $M_p$. For typical values of the parameters of a GaAs MC, vanishingly small values of $\hbar\Gamma_p$ in the range of $10^{-4}$ meV are predicted. In reality, the description of polaritons as quasi particles with a constant effective mass is inappropriate. The lower polariton branch, in particular, recovers the curvature of the bare exciton dispersion outside the strong coupling region, as seen in Fig. \ref{fig01pol}(b). The upper polariton, on the other hand, is resonant with exciton states at large momentum, to which elastic scattering can take place. In this situation the simple constant mass assumption fails \cite{Litinskaia2000,Litinskaia2001} and a microscopic model is required. Savona {\em et al.} \cite{Savona1997a} have computed the polariton broadenings by numerically solving the exciton COM Schr\"odinger equation including a disorder potential, linearly coupled to the cavity mode. The model assumed a one-dimensional system, for which disorder effects are expected to be quantitatively more relevant than in two dimensions \cite{Whittaker1998}, but was able to qualitatively reproduce the measured dependence of both upper and lower polariton inhomogeneous linewidths. The interpretation given by Savona {\em et al.} was that the polariton inhomogeneous broadening is determined by the multiple scattering processes that, starting from the initial polariton states, bring to the exciton-like states at large momentum. The smaller inhomogeneous broadening of the lower polariton branch is then explained by the energy separation between the lower polariton at $k=0$ and the exciton-like states at large $k$. For the upper polariton instead, exciton-like states of the lower polariton branch exist at the same energy, thus making multiple elastic scattering more effective. This interpretation, although technically correct, somewhat hides the actual physical origin of the broadening. 

A more elegant interpretation was proposed by D. Whittaker in two seminal papers \cite{Whittaker1998,Whittaker2000}. In simple words, the exciton-photon scattering is assumed as momentum-conserving, while disorder acts solely in the bare exciton propagation, resulting in the inhomogeneous energy distribution of the exciton spectrum. This approximation neglects all multiple scattering terms that, starting from a photon with momentum ${\bf k}$, have {\em intermediate photon states} with momentum ${\bf k}^\prime\ne{\bf k}$. The resulting Green's function of the polariton in-plane motion can be written as
\begin{equation}
G({\bf k},\omega)=\frac{1}{\hbar(\omega-\omega_p)+i\hbar\gamma_p-(\hbar\Omega/2)^2G_x({\bf k},\omega)}\,,
\label{greenpol}
\end{equation}
where the exciton Green's function $G_x({\bf k},\omega)$ enters as a self-energy contribution to the bare photon propagator.

This equation is the most complete formulation of the linear dispersion theory. It states that a momentum-conserving polariton theory including an inhomogeneous energy distribution correctly describes the polariton spectrum, provided the microscopic model of the exciton distribution is used. The smaller inhomogeneous broadening of the lower polariton branch is then simply explained in terms of the longer high-energy tail of $G_x({\bf k},\omega)$ due to exciton motional narrowing. The error with respect to a full calculation is given by the polariton multiple scattering terms, that are vanishingly small as predicted by the polariton motional narrowing argument. The linear dispersion theory was carefully checked in an experiment by Ell {\em et al.} \cite{Ell1998}, in which $G_x({\bf k},\omega)$ was independently determined by measuring the transmission spectrum of a reference QW fabricated under the same conditions as the MC sample, and applying Kramers-Kronig relations. 

As already discussed, in-plane disorder is not only responsible of a spectral broadening. The breaking of the in-plane translational symmetry still implies momentum non-conservation. Experimentally, momentum scattering manifests itself in the observation of RRS. Many experiments were carried out to study polariton RRS \cite{Gurioli2001,Hayes1998,Houdre2000,Langbein2004a,Langbein2002a,Freixanet1999,Cassabois1999}. They showed that RRS largely dominates over incoherent photoluminescence following inelastic scattering processes. The theory of polariton RRS can be developed under assuptions analogous to the linear dispersion theory, as was done by Whittaker \cite{Whittaker2000} and independently by Shchegrov {\em et al.} \cite{Shchegrov2000}. To model RRS, they considered only the first-order scattering between two polaritons described by the propagator (\ref{greenpol}), occurring via scattering of the exciton component by disorder.
This approximation is called the {\em RRS filter model} and assumes, in the same spirit as the linear dispersion theory, that cavity photons are subject to the least possible amount of momentum-changing scattering processes. The model describes well the experimental results qualitatively \cite{Whittaker2000} and, to a reasonably good extent, quantitatively \cite{Langbein2004a}.

Are the linear dispersion and RRS filter models sufficient to model all manifestations of disorder on MC polaritons? Intuitively, we expect that in MC samples with state-of-the-art interfaces -- especially at large negative cavity-exciton detuning -- the spectral density of the QW exciton is vanishingly small in correspondnce to the lower polariton energy. In such a case, the contribution from the linear dispersion or RRS filter models will eventually become comparable to the polariton multiple-scattering contribution to the line broadening. More generally, multiple scattering on disorder gives rise to Anderson localization \cite{Kramer1993}. It is therefore very important to investigate polariton localization and estimate the typical polariton localization length, that might be very large due to the light polariton effective mass. However occurring over a long range, localization still might play an important role in some aspects of polariton physics, such as the possibility of undergoing Bose-Einstein condensation, as we point out in our conclusions.

If on one side recent experiments have showed polariton ballistic propagation over several tens of microns \cite{Langbein2002c,Sermage2001a}, still signatures of polariton localization have been undoubtedly found. In particular, RRS measurements by Gurioli {\em et al.} \cite{Gurioli2001}, Houdr\'e {\em et al.} \cite{Houdre2000}, and Langbein {\em et al.} \cite{Langbein2004a} have clearly found signatures of enhanced resonant backscattering \cite{Akkermans1986,Albada1985,Kramer1993,Wolf1985}, that cannot be explained within the single-scattering picture. 
W. Langbein \cite{Langbein2004a,Langbein2002a} has also studied the polariton localization length in a high-quality MC, by measuring the momentum spread of RRS under highly directional pulsed excitation and in the long-time limit. A lower bound to the localization length in this particular sample was set to 30 $\mu$m. Further evidence of polariton localization over the $\mu$m scale was recently provided by Gurioli {\em et al.} \cite{Gurioli2005}, who characterized the angular width of the enhanced resonant backscattering peak in a MC. Their experimental data suggest that the polariton mean free path is of the order of 10 $\mu$m. Finally, in a recent photoluminescence measurement by Richard {\em et al.} \cite{Richard2005} under high excitation density and spatial resolution, the polariton population buildup in the lowest energy levels allowed a direct imaging of the shape of localized eigenstates. Their spatial extension was about 5-10 $\mu$m, consistently with the larger disorder amplitude expected in II-VI samples. To our knowledge, very few theoretical works studied polariton localization \cite{Litinskaia2001}, and the only quantitative theoretical analysis of polariton localization was carried out by Michetti {\em et al.} \cite{Michetti2005}, however using a one-dimensional model and considering Frenkel excitons in organic materials, thus characterized by a large disorder amplitude. Their conclusion is that the polaritons are localized over 1 $\mu$m in this particular system. 

Another important fact that has been well established in several experimental works, concerns the role of disorder in the cavity structure, acting on the photon component. In particular, Gurioli {\em et al.} \cite{Gurioli2001} were the first to remark that RRS occurs also in the case of large negative detuning of the cavity mode with respect to the exciton energy. In this situation, the exciton component of the involved polaritons is negligible and RRS can only be explained by assuming a disorder acting on the photon component. Furthermore, all RRS measurements \cite{Gurioli2001,Houdre2000,Langbein2004a,Langbein2002a} have brought evidence of a peculiar cross-shaped pattern, centered at the momentum of the excitation beam, appearing in the momentum plane in addition to the RRS ring. Gurioli {\em et al.} \cite{Gurioli2001}, and Langbein {\em et al.} \cite{Langbein2004a,Langbein2002a} have suggested that this is the signature of cross-hatch disorder at the interfaces of the cavity slab, originating from misfit dislocations oriented along the two orthogonal crystal axes. It is very remarkable that, despite this clear experimental evidence, all theoretical analyses of disorder effects on MC polaritons exclusively focused on excitonic disorder in the QW. Below, we extend the MC polariton model to include disorder components acting both on the exciton and on the photon degrees of freedom. By computing numerical solutions of the model equations, we review some of the main issues of polariton disorder, with special focus on the relevance of exciton and photon disorder, and briefly discuss the theoretical predictions related to the polariton RRS spectrum and localization.

\subsection{Unified model of exciton and photon disorder}

We present here a model that accounts both for excitonic and photonic components of disorder. Before, however, some considerations on the disorder correlation length must be made. In an epitaxially grown structure we expect structural disorder on a wide range of length scales. On the shortest atomic scale, disorder is caused by segregation of one material into the other and by alloy fluctuations in the case when a semiconductor alloy is used. On a longer length scale, typically of 10 to 100 nm, epitaxial growth causes the formation of monolayer-step fluctuations, whose lateral extension and amplitude can be only partially controlled by growth interruption \cite{Savona2006}. Finally, the intrinsic properties of molecular beam epitaxy can produce fluctuations on a length scale above one micron \cite{Yayon2002}. It was however shown that in QWs the long and medium range interface fluctuations are almost fully correlated between the direct and the inverse interface \cite{Savona2006}. Thus, long-range variations of the QW thickness will take place over a very long scale, practically not affecting the exciton COM motion. The medium-range component, on the other hand, produces a QW thickness fluctuation only due to the different correlation lengths of the two QW interfaces \cite{Savona2006}. Finally, the short-range component also has a contribution that is however partly averaged over the electron-hole relative wave function, as seen previously \cite{Zimmermann2003}. In addition, motional narrowing reduces the influence of medium- and short-range contributions in the white noise limit $\sigma_x/E_{cx}\ll1$ \cite{Savona1999b}. 

In a planar MC, we can define a photon effective mass as deriving from the curvature of the photon dispersion at $k=0$. In typical GaAs-based MCs, this effective mass is as small as $3\times10^{-5}$ times the free electron mass. We will see below that Maxwell equations for the photon motion along the plane are equivalent to an effective Schr\"odinger equation where the photon effective mass enters the kinetic term. Hence, similar considerations as for the exciton COM motion apply. Given the very light mass, disorder components with correlation length $\xi_p\le 100$ nm have $E_{cp}\ge 125$ meV. A realistic upper bound for the energy fluctuation of the photon mode is $\sigma_{p}\le 0.5$ meV, deduced from the measured cavity-mode linewidth in modern samples. The same scaling argument used for the exciton suggests that short- and medium-range components of the MC disorder scarcely affect the in-plane motion of the photon. The long-range disorder component however is still relevant. In particular, given the large thickness of the cavity slab, the fluctuations characterizing the two interfaces are expected to be uncorrelated, differently from a QW, resulting in an overall fluctuation of the cavity thickness that will locally affect the resonant frequency of the cavity mode.

Since the cavity-slab thickness varies very smoothly over lengths comparable to the wavelength, we make the simplifying assumption that the electromagnetic field inside the cavity is locally equal to that of an ideally planar cavity. We introduce the further approximation of assuming a scalar electric field, thus neglecting the possible polarization vectors of the field and the corresponding selection rules for the linear coupling to the exciton spin states \cite{Andreani1994a,Tassone1992}. The interplay between disorder and the vector nature of the polariton field was recently investigated both theoretically \cite{Kavokin2005} and experimentally \cite{Langbein2002c}, and results in a rotation of the polarization direction of the RRS field over a timescale of a few tens of ps. Within our scalar approximation, the electric field at position ${\bf r}=({\bf R},z)$ is
\begin{equation}
{E}({\bf r})={E}({{\bf R}})\exp(ik_z({{\bf R}})z)\,,
\label{ansatz}
\end{equation}
where $k_z({{\bf R}})$ is a position-dependent photon momentum along the $z$-direction. 
By neglecting terms proportional to the gradient of $k_z({{\bf R}})$, Maxwell equations give
\begin{equation}
\nabla^2_{\rho}{E}({{\bf R}})+\left(\frac{\omega^2}{c^2}\epsilon_0-k_z^2({{\bf R}})\right){E}({{\bf R}}) + 4\pi\frac{\omega^2}{c^2}{P}({{\bf R}})=0\,,
\label{maxwell2d}
\end{equation}
where $\epsilon_0$ is the spacer dielectric constant, while ${P}({{\bf R}})$ is the excitonic macroscopic polarization field. If we consider energies close to the photon-mode energy at zero momentum $\hbar\omega_p$, we can expand the $\omega^2$ term in (\ref{maxwell2d}) and keep only the linear term in the expansion. The equation can now be Fourier transformed to the time domain, resulting in a Schr\"odinger-like equation for the electric field.
We can expand $k_z^2({{\bf R}})$ to the first order in the fluctuations
\begin{equation}
k_z^2({{\bf R}})\approx k_p^2+2k_p\delta k_z({{\bf R}})\,,
\end{equation}
where $k_p=\sqrt{\epsilon_0}\omega_p/c$. We further define the photon effective mass $M_{p}$ through the kinetic term in the equation, as
\begin{equation}
\frac{\hbar c^2}{\omega_p\epsilon_0}=\frac{\hbar^2}{2M_{p}}
\end{equation}
We finally obtain
\begin{equation}
i\hbar\frac{\partial}{\partial t}{E}({{\bf R}},t)=
-\frac{\hbar^2}{2M_{p}}\nabla^2_{\rho}{E}({{\bf R}},t)+V_p({{\bf R}}){E}({{\bf R}},t) - 2\pi\frac{\hbar\omega_p}{\epsilon_0}{P}({{\bf R}},t)\,,
\label{MaxwellSchr}
\end{equation}
where we have defined the effective disorder potential affecting the in-plane motion of the photon
\begin{eqnarray}
V_p({{\bf R}})&=&\frac{\hbar c}{\sqrt{\epsilon_0}}\delta k_z({{\bf R}})\nonumber\\
&=&-\hbar\omega_p\frac{\delta L_c({{\bf R}})}{\lambda_0+L_{DBR}}\,.
\label{photpot}
\end{eqnarray}
Here, we have further related the potential energy fluctuations to the cavity thickness fluctuations $\delta L_c({{\bf R}})$ through the cavity thickness $\lambda_0$ and the effective DBR penetration length $L_{DBR}$ \cite{Macleod2001}. 

The exciton macroscopic polarization $P({{\bf R}},t)$ depends linearly on the exciton center-of-mass wavefunction $\psi({{\bf R}},t)$. This latter is governed by a Schr\"odinger equation that includes the disorder potential acting on the QW exciton and the electric field as an external source \cite{Savona1999b,Zimmermann2003}. If we approximate the linear exciton-photon coupling by a constant factor $\hbar\Omega/2$, which holds for momentum components smaller than $k_p$, this finally leads to two coupled Schr\"odinger equations that we can rewrite as
\begin{eqnarray}
i\hbar\frac{\partial}{\partial t}{E}({{\bf R}},t)&=&\hbar(\omega_p-i\gamma_p){E}({{\bf R}},t)\nonumber\\
&-&\frac{\hbar^2}{2M_{p}}\nabla^2_{\rho}{E}({{\bf R}},t)+V_p({{\bf R}}){E}({{\bf R}},t) + \frac{\hbar\Omega}{2}\psi({{\bf R}},t)\,,
\label{pol1}\\
i\hbar\frac{\partial}{\partial t}{\psi}({{\bf R}},t)&=&\hbar(\omega_x-i\gamma_x){\psi}({{\bf R}},t)\nonumber\\
&-&\frac{\hbar^2}{2M_{x}}\nabla^2_{\rho}{\psi}({{\bf R}},t)+V_x({{\bf R}}){\psi}({{\bf R}},t) + \frac{\hbar\Omega}{2}{E}({{\bf R}},t)\,,
\label{pol2}
\end{eqnarray}
where $\hbar\omega_x$ is the ground exciton energy and we have introduced the damping rates $\gamma_p$ and $\gamma_x$, for the photon and exciton respectively. These rates describe phenomenologically the photon escape rate from the MC and the non-radiative exciton lifetime.

Equations (\ref{pol1}) and (\ref{pol2}) model the polariton dynamics in presence of disorder potentials for both the exciton and the photon fields. The vacuum-field Rabi splitting is given by $\hbar\Omega$ and is used as an input parameter in the calculations. The main difficulty in treating Eqs. (\ref{pol1}) and (\ref{pol2}) is related to the large difference in the correlation length of the disorder potentials acting on the exciton and photon components. A basic requirement for the numerical solution of a time-dependent Schr\"odinger equation in real space, is that the size of the simulation grid step $\Delta$ is small enough to correctly sample the features of the disorder potential. In more rigorous terms, the criterion is that the hopping energy $T=\hbar^2/(2M\Delta^2)$ be much larger than the two energy parameters $E_c$ and $\sigma$, characterizing respectively the correlation length and amplitude of the disorder potential. In the present problem, however, these two parameters take different values for the exciton and the photon. While the two problems are likely to be characterized by comparable values of $\sigma_x$, and $\sigma_p$, the values of $E_{cx}$ and $E_{cp}$ will differ considerably, due to the large difference in effective mass and correlation length. In particular, $E_{cx}$ is much larger than  $E_{cp}$. Then, in order to correctly model the exciton kinetics, one needs to choose a value of $\Delta$ reasonably smaller than 10 nm. At the same time, the simulation domain must extend over several $\mu$m in order to model the photon propagation. This situation calls for large scale simulations with thousands of grid points for each spatial dimension \cite{Whittaker1998}. A possible way of overcoming this limitation is given by the remark that, for polaritons in the strong coupling region, the in-plane kinetics is governed essentially by the photon effective mass. Hence, taking the limit $M_x\rightarrow\infty$ should leave the results in the strong coupling region unchanged, while affecting the predictivity of the model in the exciton-like part of the lower polariton branch. We call this approximation {\em local oscillator model}, as the exciton part of the dynamics is now described by a set of independent local oscillators at each simulation site. In the absence of photon disorder, the validity of this approximation was suggested by Whittaker \cite{Whittaker1998}, who showed that the influence of exciton disorder on the polaritons in the strong coupling region is to all extents negligible, already in the case $\sigma_x\lesssim\omega$.

In order to test the local oscillator model also in presence of photon disorder, we have carried out large-scale numerical simulations of the polariton spectrum, comparing the local-oscillator and full kinetic results. A simulation domain of $20\times20~\mu\mbox{m}^2$ was sampled on a $2048\times2048$ grid. Disorder potentials were assumed Gauss-correlated in space, with $\sigma_x=0.5$ meV, $\xi_x=10$ nm, $\sigma_p=0.3$ meV, and $\xi_p=1~\mu$m, corresponding to currently available good quality systems. The Rabi splitting was $\hbar\Omega=3.8$ meV, as for a typical sample embedding a single GaAs QW, while zero exciton-cavity detuning was assumed. Equations (\ref{pol1}) and (\ref{pol2}) were solved using the kernel-polynomial method \cite{Weisse2006}. This method gives direct access to the spectrally resolved Green's function. With respect to time-evolution methods \cite{Whittaker1998} it has the advantage of a rapid convergence, as the relative energy resolution scales as $N_m^{-3/2}$ with the number $N_m$ of momenta in the polynomial expansion, at the boundary of the $8T$-wide tight-binding spectral domain. We needed $N_m=160000$ to reach an energy resolution better than the typical polariton homogeneous broadening of about $50~\mu$eV. As each momentum requires one application of the linear operator defined by the right-hand side of (\ref{pol1})-(\ref{pol2}), this number should be compared to the number of time-steps in a time-evolution simulation which is typically of a few millions for the same energy resolution. As initial state, we assumed an excitation pulse of 1 $\mu$m Gauss-width. The simulation gives access to the polariton spectral function, defined as the Fourier-transformed electric field as $I_{\bf k}(\omega)=|E_{\bf k}(\omega)|^2$. It corresponds to the field intensity that would be measured in a RRS measurement.

\begin{figure}
\centerline{\includegraphics[width=5.0in]{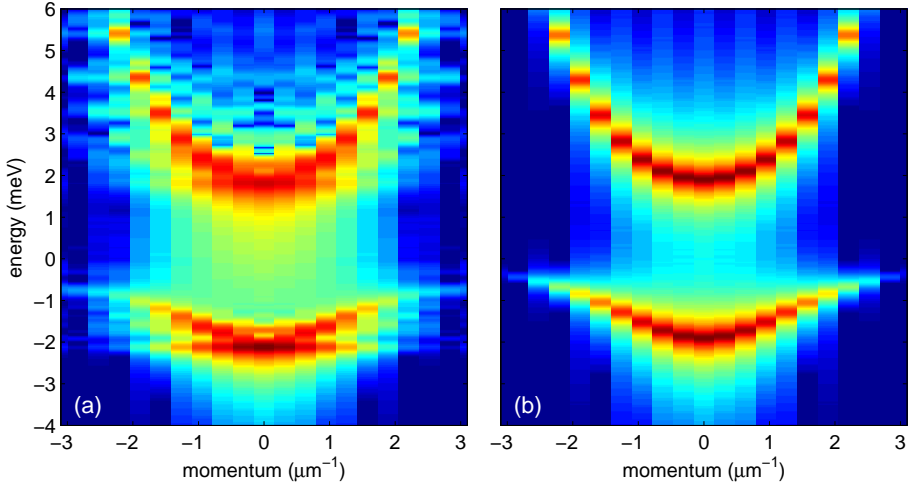}}
\caption{(a): Simulated polariton spectral function for a single realization of photon and exciton disorder potentials (parameters as in the text). (b): Same as in (a) but without disorder. Logarithmic scale over 7 decades from blue to red.}
\label{fig03pol} 
\end{figure}

Fig. \ref{fig03pol}(a) displays the simulated energy-momentum spectral function using a logarythmic-intensity color scale, for a single disorder realization and within the full kinetic model. For comparison, in Fig. \ref{fig03pol} (b) we display the spectrum in the absence of disorder on both components. Both plots clearly show the coarse momentum grid and the corresponding energy discretization, which are due to having chosen a simulation area of only $20\times20~\mu\mbox{m}^2$. The ideal spectrum in Fig. \ref{fig03pol} (b) is characterized by a single Lorentzian resonance for each polariton branch. On the other hand, speckles are clearly visible in the spectrum of the disordered system in Fig. \ref{fig03pol} (a). Although the log-scale plot does not provide a clear quantitative information, still two important features can be guessed. First, both upper and lower polaritons display a flat pattern at the bottom of the respective bands, extending over about $\pm1~\mu\mbox{m}^{-1}$. It is a spectral signature of polariton localization. The numerical approach that we use does not give direct access to the polariton wave functions, but we can guess that in this disorder realization at least one well localized state of both upper and lower polariton exist. The localization length can be inferred from the extension on the momentum axis and amounts to a few $\mu$m. The second important feature to be noticed is the rather sharp spectrum of the lower branch at large momentum, where the polariton is almost fully exciton-like. Here, $\sigma_x/E_{cx}\approx0.5$ and motional narrowing results in an exciton inhomogeneous broadening significantly smaller than the potential energy broadening $\sigma_x$.

\begin{figure}
\centerline{\includegraphics[width=5.0in]{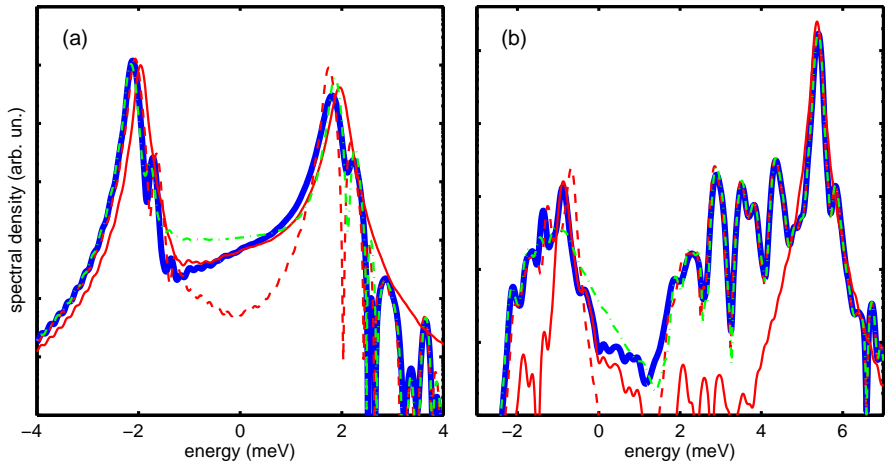}}
\caption{(a): Simulated polariton spectral function at ${\bf k}=0$ for a single disorder realization. Thick (blue) line: $\sigma_x=0.5$ and $\sigma_p=0.3$ meV. Thin (red) line $\sigma_x=0$ and $\sigma_p=0.3$ meV. Dashed (red) line: $\sigma_x=0.5$ meV and $\sigma_p=0$. Dot-dashed (green) line: local oscillator model. (b): Same as in (a) computed at ${\bf k}=(2.2,0)~\mu\mbox{m}^{-1}$.}
\label{fig04pol} 
\end{figure}

In order to gain a better insight into these numerical results, we plot in Fig. \ref{fig04pol} (a) and (b) the polariton spectrum taken at $k=0$ and $k=2.2~\mu\mbox{m}^{-1}$ respectively. In each plot, four curves are compared, corresponding (i) to the full simulation, (ii) to setting $\sigma_x=0$ and $\sigma_p=0.3$ meV, (iii) to setting $\sigma_x=0.5$ meV and $\sigma_p=0$, and (iv) to the local oscillator limit in which both disorder amplitudes are nonzero. At $k=0$, the spectrum obtained by setting $\sigma_x=0$ matches very well to the full simulation. In particular, both the linewidths and the relative intensity of upper and lower polaritons are well reproduced. We notice a small energy shift due to the overall exciton localization energy and some speckles that are not present in the curve with $\sigma_x=0$. This first comparison suggests that exciton disorder does affect the polariton spectrum, but only marginally, whereas the main spectral features are essentially determined by photon disorder. Indeed, the spectrum obtained by setting $\sigma_p=0$ differs more significantly from the full simulation. It underestimates in particular the upper polariton linewidth while largely overestimating its peak intensity. Finally, we notice that the local oscillator limit reproduces fairly well the full result. Again, the upper polariton strength is overestimated, although less dramatically than in the spectrum with $\sigma_p=0$. This discrepancy is expected from the local oscillator model where the bare exciton spectrum is fully symmetric around $\omega=0$. We now turn to Fig. \ref{fig04pol} (b), corresponding to the spectrum at $k_x=2.2~\mu\mbox{m}^{-1}$. Here, contrarily to the previous case, the spectrum with $\sigma_x=0$ differs significantly from the full simulation, which is instead almost exactly reproduced by the simulation with $\sigma_p=0$. This difference is expected, as in this region photon and exciton modes are practically uncoupled. In particular, the lower branch is almost fully exciton-like and disorder on the exciton component plays a major role. Finally, we point out that the local-oscillator limit in Fig. \ref{fig04pol} (b) completely misses the sharp features in the spectral region of the lower polariton. The reason is again clear, as in this region the exciton motional narrowing due to its finite mass is effective in determining the spectral features. In conclusion, the comparison between the four different cases gives two important indications. First, in the strong-coupling region the main spectral features are determined almost exclusively by the disorder acting on the photon component, whereas both disorder components become important in the regions where exciton and photon modes are decoupled. Second, the local oscillator limit is a very good approximation in the strong coupling region, where exciton motional narrowing does not affect the polariton spectrum \cite{Litinskaia2001,Whittaker1998}.

\begin{figure}
\centerline{\includegraphics[width=4.0in]{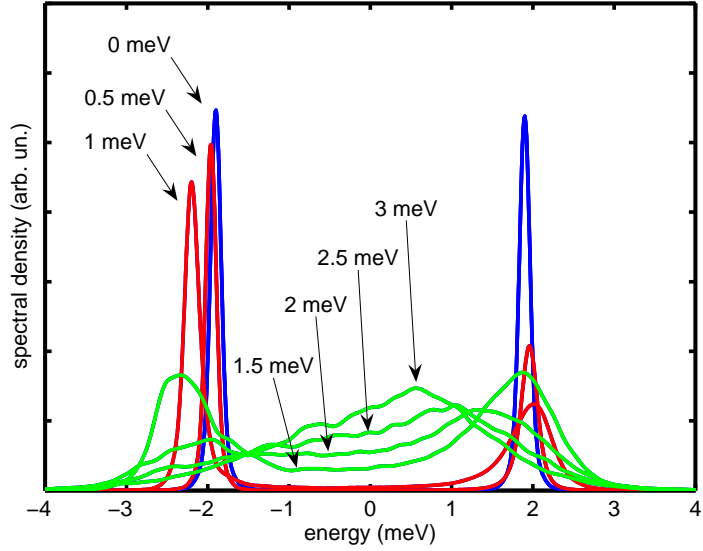}}
\caption{Simulated polariton spectral function at ${\bf k}=0$ for a single exciton disorder realization, with $\sigma_p=0$ and $\sigma_x$ taking the values indicated in the plot.}
\label{fig05pol} 
\end{figure}

To conclude this section, we take advantage of the simulations to study how the exciton disorder affects the polariton spectrum for increasing disorder amplitude. In Fig. \ref{fig05pol}, we display polariton spectra computed at $k=0$ for increasing values of the exciton disorder amplitude $\sigma_x$, while keeping $\sigma_p=0$. The polariton linewidth is practically unaffected by exciton disorder for $\sigma_x<1$ meV. In particular, the lower polariton branch proves to be more robust than the upper one to inhomogeneous broadening. Starting from $\sigma_x=1.5$ meV, both polariton peaks display a considerable inhomogeneous broadening of the same order as $\sigma_x$. Eventually, polariton localization becomes the dominant mechanism and the Rabi splitting is completely washed out for $\sigma_x>2$ meV. The behaviour observed in this numerical simulation essentially confirms the result that were obtained by D. Whittaker \cite{Whittaker1998}. The conclusion of that work was that the polariton inhomogeneous broadening originates almost only from multiple scattering to more or less localized exciton states weakly coupled to the photon mode. For small disorder, these states have a vanishing spectral density at the energy of the lower polariton branch, thus explaining the robustness of this line to inhomogeneous broadening. Multiple scattering of polaritons within the strong coupling region contributes only negligibly, due to the very short spatial range of the QW disorder, thus also preventing polariton localization. Here, by considering an additional source of disorder acting on the photon component, we might generalize this statement as follows. {\em Polariton multiple scattering within the strong coupling region -- and consequent polariton localization -- are exclusively due to scattering by the long-range disorder acting on the photon component, provided the amplitude of exciton disorder is significantly smaller than the vacuum-field Rabi splitting}. Consistently, the inhomogeneous broadening of the lower polariton line is mostly due to photon disorder.

\subsection{Modeling polariton RRS and localization}

In the previous section, we have checked the validity of the local oscillator approximation. We can now take advantage of this approximation for performing numerical simulations of large spatial domains. For the numerical simulations, we remind that the step $\Delta$ of the simulation grid must be chosen in such a way that the hopping energy $T=\hbar^2/(2M\Delta^2)$ is much larger than any other characteristic energy of the system. Contrarily to the full simulation, now we are only limited by the very small photon effective mass $M_p$, which results in very large values of $T$. This makes it possible to use much larger values of the grid step, therefore giving access to simulations over a very large spatial domain.

As an example, we now perform numerical simulations of the polariton optical response within the local oscillator approximation, on a simulation domain of $400\times400~\mu\mbox{m}^2$ sampled on a $512\times512$ grid. The step $\Delta\approx780$ nm is now large enough to make time-integration simulations more advantageous than the kernel-polynomial method, as they give simultaneous access to the RRS amplitude at all values of the in-plane momentum. We compare two simulations, both assuming an exciton disorder amplitude $\sigma_x=0.2$ meV, while the photon disorder amplitude was chosen as $\sigma_p=0.2$ meV and $\sigma_p=0$ respectively. The photon disorder was assumed Gauss correlated in space, with a correlation length $\xi_p=2~\mu$m. The Rabi splitting was again $\hbar\Omega=3.8$ meV. For all simulations, the initial excitation is provided by a Gauss pulse of 100 fs duration and 1 $\mu$m diameter, centered at $\omega=0$. In this way, the Fourier-transformed electric field amplitude $E_{\bf k}(\omega)$ models the RRS field as a function of momentum and frequency. 

\begin{figure}
\centerline{\includegraphics[width=5.0in]{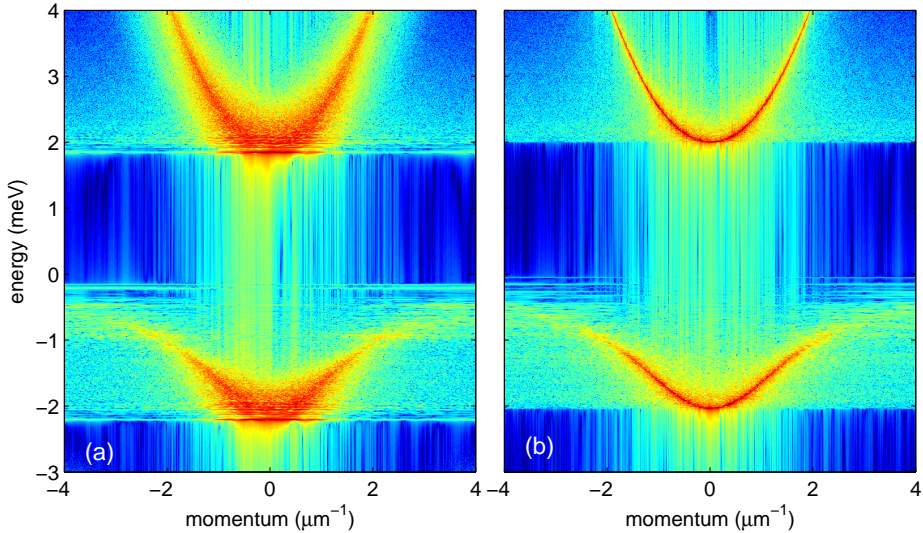}}
\caption{(a): Simulated polariton RRS intensity for a single realization of photon and exciton disorder potentials having $\sigma_x=0.2$ meV and $\sigma_p=0.2$ meV (for the other parameters see text). (b): Same as in (a) but with $\sigma_p=0$. For both plots, a logarithmic scale over 5 decades from blue to red is used. (c) Simulated polariton localization length as a function of energy.}
\label{fig06pol} 
\end{figure}

Figs. \ref{fig06pol} (a) and (b) show the simulated RRS intensity as a function of in-plane momentum and energy. In Fig. \ref{fig06pol} (a) the case with $\sigma_x=\sigma_p=0.2$ meV is displayed, while panel (b) displays the case with $\sigma_x=0.2$ meV and $\sigma_p=0$. Both images are characterized by a fine speckle pattern resulting from the exciton disorder component. In panel (a), the spectral broadening of both lower and upper polariton in the strong coupling region is however significantly larger than in panel (b), as a result of the photon disorder. This confirms what already observed in the previous section, namely that polaritons are almost exclusively affected by disorder of the cavity structure, whereas QW disorder has mainly influence on the exciton-like part of the dispersion only. By comparing Figs. \ref{fig06pol} (a) and (b), we also notice that the first spectrum is rigidly shifted to slightly lower energy with respect to the second one. Correspondingly, in panel (a) we also notice the presence of several sharp spectral features, forming a flat energy-momentum pattern, at the bottom of both the upper and the lower polariton branches, that are instead absent in panel (b). As already pointed out in the previous section, both are clear signatures of polariton spatial localization due to the long-range photon disorder. In particular, the flat features in the energy-momentum RRS plots correspond to discrete polariton eigenstates that are localized via their photon component. The rigid energy shift corresponds to the localization energy in the specific realization of the photon disorder used for the simulation. Apart from the much better energy and momentum resolution and finer speckle pattern, the only difference between Fig. \ref{fig06pol} (a) and Fig. \ref{fig03pol} (a) resides in the exciton-like part of the lower polariton branch, whose spectral broadening is overestimated by the local oscillator approximation.

The simulation over such a large spatial domain will make it possible to evaluate the polariton localization length under different assumptions for the exciton and photon components of disorder. This analysis is however beyond the scope of the present review. For the parameters used in the previous example, we can estimate the localization length from the momentum spread of the RRS spectral pattern. In the case with nonzero photon disorder amplitude, the flat features at the bottom of both polariton branches correspond to a localization length of about 10 $\mu$m, in agreement with experimental estimates on a GaAs/AlAs sample with comparable interface quality \cite{Langbein2002a,Richard2005}. The case with only exciton disorder, on the other hand, gives 50 $\mu$m as a strict lower bound for the polariton localization length at the band bottom. 

\section{Conclusions}

We have presented an overview of the research on disorder effects on excitons in QWs and polaritons in MCs. As a general rule, the most direct consequences of disorder are the inhomogeneous spectral broadening and the RRS in the optical response. Microscopically, disorder in two dimensions is expected to give rise to spatial localization of the COM motion. In the case of QW excitons, COM localization was the object of a careful analysis, both theoretical and experimental. The effect of disorder on polaritons, on the other hand, awaits a more detailed anaysis. The existing studies have mainly focused on the effect of QW disorder. The effect of disorder in the dielectric structure of the MC was pointed out by some recent experimental studies but never investigated in detail. Moreover, due to the very light effective mass of the polariton quasi-particle at the bottom of the polariton dispersion, the idea that localized states of polaritons might arise was overlooked by most studies.

In the second part of this work, we have presented an original study of disorder on polaritons, based on a model that accounts for QW and MC disorder components on equal footing. Preliminary numerical results suggest that polaritons are mostly affected by the long-range photon component of disorder, which is responsible for RRS and polariton localization. The localization length of polaritons, predicted for typical parameters of today's semiconductor heterostructures, ranges from a few $\mu$m to a few tens of $\mu$m at the bottom of the polariton band, in agreement with the few existing experimental studies.

Localization is very important in view of the recent experiments aimed at achieving Bose-Einstein condensation of QW excitons \cite{Snoke2002,Gorbunov2006,Butov2004,Voros2006} or MC polaritons \cite{Deng2003,Kasprzak2006} in two-dimensional systems. In a Bose gas in two dimensions, conventional Bose-Einstein condensation with occurrence of off-diagonal long-range order cannot take place, as stated by the Hohenberg-Mermin-Wagner theorem \cite{Hohenber1967}. Theory instead predicts the Berezinskii-Kosterlitz-Thouless transition \cite{Kosterlitz1973,Berezinski1972} from a disordered phase with spontaneous formation of vortices, to an ordered phase called quasi-condensate, characterized by a Bogolubov spectrum of collective excitations and a polynomial decay of long-range spatial correlations \cite{Pitaevskii2003}. If however the Bose gas is trapped \cite{Pitaevskii2003,Lauwers2003,Petrov2000,Ketterle1996,Bagnato1991} or subject to disorder \cite{Lenoble2004}, then a genuine Bose-Einstein condensate can be expected, provided the localization occurs over a sufficiently small scale. With increasing size of the condensate, on the other hand, a crossover from genuine Bose-Einstein condensation to a Berezinskii-Kosterlitz-Thouless behaviour should take place. A thorough characterization of exciton and polariton localization and of its influence on the phase transition is then a necessary step for understanding the nature of these collective phenomena.

\ack
The first part of this article, dealing with disorder and exciton localization in QWs, is mostly derived from the research carried out over the last fifteen years by Roland Zimmermann and Erich Runge \cite{Zimmermann2003}, to whom I am also indebted for the many invaluable discussions and the fruitful collaboration. I am also particularly grateful to Wolfgang Langbein for pointing out many experimental facts on MC-polaritons. I acknowledge many interesting discussions with Beno\^it Deveaud, Antonio Quattropani and Paolo Schwendimann. This work was supported by the Swiss National Science Foundation through project N. 620-066060.

\section*{References}


\end{document}